\documentclass[runningheads]{llncs}
\usepackage[T1]{fontenc}
\usepackage{graphicx}
\usepackage{hyperref}
\usepackage{color}

\usepackage[a4paper, margin=1.5in]{geometry}
\begin{document}
\title{The Sketchfab 3D Creative Commons Collection (S3D3C)}
%
%
\author{Florian Spiess\inst{1}\orcidID{0000-0002-3396-1516} \and
Raphael Waltenspül\inst{1}\orcidID{0009-0004-9622-7265} \and
Heiko Schuldt\inst{1}\orcidID{0000-0001-9865-6371}}
\authorrunning{F. Spiess et al.}
%
\institute{University of Basel, Basel, Switzerland \email{first.last@unibas.ch}}
\maketitle              
\begin{abstract}
  The technology to capture, create, and use three-dimensional (3D) models has become increasingly accessible in recent years.
  With increasing numbers of use cases for 3D models and collections of rapidly increasing size, better methods to analyze the content of 3D models are required.
  While previously proposed 3D model collections for research purposes exist, these often contain only untextured geometry and are typically designed for a specific application, which limits their use in quantitative evaluations of modern 3D model analysis methods.
  
  In this paper, we introduce the Sketchfab 3D Creative Commons Collection (S3D3C), a new 3D model research collection consisting of 40,802 creative commons licensed models downloaded from the 3D model platform Sketchfab.
  By including popular freely available models with a wide variety of technical properties, such as textures, materials, and animations, we enable its use in the evaluation of state-of-the-art geometry-based and view-based 3D model analysis and retrieval techniques.
  \keywords{3D Model Dataset \and Sketchfab \and Creative Commons.}
\end{abstract}
\section{Introduction}

The importance of three-dimensional (3D) models is increasing in many industries.
To develop and improve methods and tools for the effective analysis and use of these models, 3D model collections are required to evaluate and compare performance.
A number of 3D model datasets have been made available over the previous decades.
However, these datasets are usually collected for a very specific task and are often limited in number, variety, and/or quality of annotation.
To evaluate modern, state-of-the-art 3D model research applications, such as content-based 3D model analysis and retrieval, a research collection is needed that contains not only a large number of 3D models, but also a variety of models representing the breadth of 3D models in use today, including examples of models containing textures, materials, animations, and even audio.

In this paper we introduce the Sketchfab 3D Creative Commons Collection (S3D3C), a new 3D model collection for use in research applications.
The 40,802 models are available on Sketchfab\footnote{\url{https://sketchfab.com}} under creative commons licenses, contain both digitally created models as well as 3D scans, belong to a wide range of categories ranging from cultural heritage to food, include textured and animated models, and have associated user provided metadata, including title, description, category, and tags.

In the remainder of this paper we briefly introduce existing 3D model datasets in \autoref{sec:existing-collections}, describe and analyze the Sketchfab 3D Creative Commons Collection in \autoref{sec:s3d3c}, discuss use-cases, limitations, and future work in \autoref{sec:discussion}, and conclude in \autoref{sec:conclusion}.

\section{Existing 3D Model Collections}
\label{sec:existing-collections}

Some 3D model research collections have already been proposed.
In this section, we briefly summarize the most common collections.

The Princeton Shape Benchmark~\cite{shilanePrincetonShapeBenchmark2004} contains 1,814 polygonal 3D models.
These models are labelled with 92 hierarchical classes.
ModelNet~\cite{wu3DShapeNetsDeep2015} contains 151,128 3D CAD models, containing only geometry, labelled with 660 categories.
ShapeNet~\cite{changShapeNetInformationRich3D2015} contains 3,000,000 raw models, which are not publicly available.
The largest publicly available subset, ShapeNetCore, contains 51,300 3D objects categorized into 55 common object categories.

While most of these collections contain large numbers of 3D models and all of those introduced include class labels, the primary focus for the majority is on model shapes.
In existing collections, little emphasis is placed on modern technical properties of 3D models, such as advanced materials, textures, and animations.
In addition most previously proposed collections contain very little metadata, which is usually limited to an associated class.

\section{The Sketchfab 3D Creative Commons Collection}
\label{sec:s3d3c}

Our proposed 3D model collection contains 40,802 models collected from the 3D model platform Sketchfab.
In this section we present the process through which we selected and collected the models, how the collection can be downloaded by peers, and an initial analysis of the models within the collection.

\subsection{Collection Process}

To collect the 3D models from Sketchfab, we used the Sketchfab data\footnote{\url{https://sketchfab.com/developers/data-api/v3}} and download\footnote{\url{https://sketchfab.com/developers/download-api}} APIs.
We selected the 3D models for the collection using the \texttt{search} endpoint of the data API.
Using the license filter argument, we selected only creative commons licensed models.
In order to include the highest quality models, we selected those with the highest user ranking for each license type.
High ranking models have been reviewed and liked by other users, which we expect to be a good indicator of model quality.

In addition to the models, which we downloaded as glTF\footnote{\url{https://www.khronos.org/gltf}}, we also downloaded the model information provided by Sketchfab as metadata.
The model metadata were collected between May 2023 and January 2024, with the majority of the metadata having been collected in January 2024.
The statistics we present in the following section have been calculated using the metadata from the models for which we were able to successfully download both the model data and metadata.

The metadata including the UIDs required to download the models from Sketchfab are available on the S3D3C OSF project\footnote{\url{https://osf.io/w2qu9}}.
We plan to also make the models available for download directly to ensure that models removed from Sketchfab remain available.

\subsection{Model Statistics}

Using the downloaded model metadata provided by Sketchfab, we calculate statistics on our proposed collection.
We consider particularly the technical aspects of the models usually not included or reported for previously proposed 3D model collections, such as which models contain texture files, animations, and sound.
A summary of the most important model statistics is shown in \autoref{tab:statistics}.

\begin{table*}
    \centering
    \caption{
        Model statistics in total and by license.
        A model is counted for having a texture file, animation, or sound if the Sketchfab metadata reports a respective count greater than zero.
        Has tag/category count models with at least one respectively.
    }
    \begin{tabular*}{\linewidth}{@{\extracolsep{\fill}}l|rrrrrrrrrr}
        & \multicolumn{2}{c}{\textbf{CC0}}   & \multicolumn{2}{c}{\textbf{CC-BY}}   & \multicolumn{2}{c}{\textbf{CC-BY-SA}}   & \multicolumn{2}{c}{\textbf{CC-BY-ND}}   & \multicolumn{2}{c}{\textbf{Total}} \\\hline
        Models & 3,858 &  & 14,819 &  & 13,457 &  & 8,668 &  & 40,802 &  \\
        Texture file & 3,150 & 81.6\% & 13,080 & 88.3\% & 9,927 & 73.8\% & 5,176 & 59.7\% & 31,333 & 76.8\% \\
        Animation & 67 & 1.7\% & 1,612 & 10.9\% & 441 & 3.3\% & 417 & 4.8\% & 2,537 & 6.2\% \\
        Sound & 10 & 0.3\% & 279 & 1.9\% & 81 & 0.6\% & 9 & 0.1\% & 379 & 0.9\% \\
        Has description & 3,712 & 96.2\% & 13,428 & 90.6\% & 9,790 & 72.8\% & 4,794 & 55.3\% & 31,724 & 77.8\% \\
        Has tag & 3,440 & 89.2\% & 14,139 & 95.4\% & 12,152 & 90.3\% & 7,081 & 81.7\% & 36,812 & 90.2\% \\
        Has category & 3,570 & 92.5\% & 13,602 & 91.8\% & 8,651 & 64.3\% & 4,909 & 56.6\% & 30,732 & 75.3\% \\
        Age restricted & 0 & 0.0\% & 98 & 0.7\% & 0 & 0.0\% & 0 & 0.0\% & 98 & 0.2\% \\
    \end{tabular*}
    \label{tab:statistics}
\end{table*}

It is particularly noteworthy, that $6.2\%$ of the models in the collection contain animation and $0.9\%$ contain sound.
While these are only small parts of the collection, the inclusion of these models increases the variety of model modalities.

\begin{table}
    \centering
    \caption{Models per category for the 18 Sketchfab categories.}
    \begin{tabular}{l|rr}
        \textbf{Category} & \textbf{Count} &  \\\hline
        Cultural Heritage \& History & 7,060 & 17.3\% \\
        Science \& Technology & 4,597 & 11.3\% \\
        Characters \& Creatures & 4,589 & 11.2\% \\
        Architecture & 4,234 & 10.4\% \\
        Art \& Abstract & 4,161 & 10.2\% \\
        Furniture \& Home & 3,617 & 8.9\% \\
        Animals \& Pets & 3,007 & 7.4\% \\
        Weapons \& Military & 2,570 & 6.3\% \\
        Nature \& Plants & 2,562 & 6.3\% \\
        Cars \& Vehicles & 2,213 & 5.4\% \\
        Electronics \& Gadgets & 1,978 & 4.8\% \\
        Places \& Travel & 1,955 & 4.8\% \\
        Fashion \& Style & 1,097 & 2.7\% \\
        People & 903 & 2.2\% \\
        Food \& Drink & 899 & 2.2\% \\
        Sports \& Fitness & 214 & 0.5\% \\
        Music & 163 & 0.4\% \\
        News \& Politics & 55 & 0.1\% \\
    \end{tabular}
    \label{tab:categories}
\end{table}

\begin{table}
    \centering
    \caption{Models per tag for the top 20 tags.}
    \begin{tabular}{l|rr}
        \textbf{Tag} & \textbf{Count} &  \\\hline
        qlone & 4,803 & 11.8\% \\
        tiltbrush & 3,013 & 7.4\% \\
        blender & 2,630 & 6.4\% \\
        noai & 2,563 & 6.3\% \\
        free & 2,218 & 5.4\% \\
        lowpoly & 2,178 & 5.3\% \\
        substancepainter & 1,798 & 4.4\% \\
        3dlivescanner & 1,550 & 3.8\% \\
        photogrammetry & 1,529 & 3.7\% \\
        substance & 1,365 & 3.3\% \\
        game & 1,320 & 3.2\% \\
        openbrush & 1,191 & 2.9\% \\
        low-poly & 1,034 & 2.5\% \\
        gameready & 992 & 2.4\% \\
        lidar & 918 & 2.2\% \\
        low & 913 & 2.2\% \\
        gameasset & 910 & 2.2\% \\
        old & 909 & 2.2\% \\
        prop & 871 & 2.1\% \\
        poly & 870 & 2.1\% \\
    \end{tabular}
    \label{tab:tags}
\end{table}

With regards to available metadata, $75.3\%$ of the models have at least one category associated with them.
Sketchfab models can be placed in up to two of the 18 fixed categories supported by Sketchfab.
A more detailed distribution of the number of models associated with each of these 18 categories is shown in \autoref{tab:categories}.
The most common category associated with models in the collection is ``Cultural Heritage \& History'' with 7,060 associated models, which make up 17.3\% of the S3D3C.

Out of all models in the collection, $90.2\%$ are associated with at least one tag.
In contrast to the fixed categories defined by Sketchfab, tags can be freely defined by the user.
\autoref{tab:tags} shows the number of models tagged with each of the 20 most commonly used tags in the collection.
The most used tags are mainly concerned with the tool used to create the models, such as ``qlone'', ``tiltbrush'', and ``blender''.
Tags that are more descriptive of the model content, such as ``lowpoly'' and ``old'', appear less frequently.

\section{Discussion}
\label{sec:discussion}

In this section, we discuss some favorable properties of the Sketchfab 3D Creative Commons Collection in the context of examples from the collection, and reflect on limitations of the collection.

\subsection{Properties and Examples}

\begin{figure}
    \centering
    \includegraphics[width=.5\linewidth]{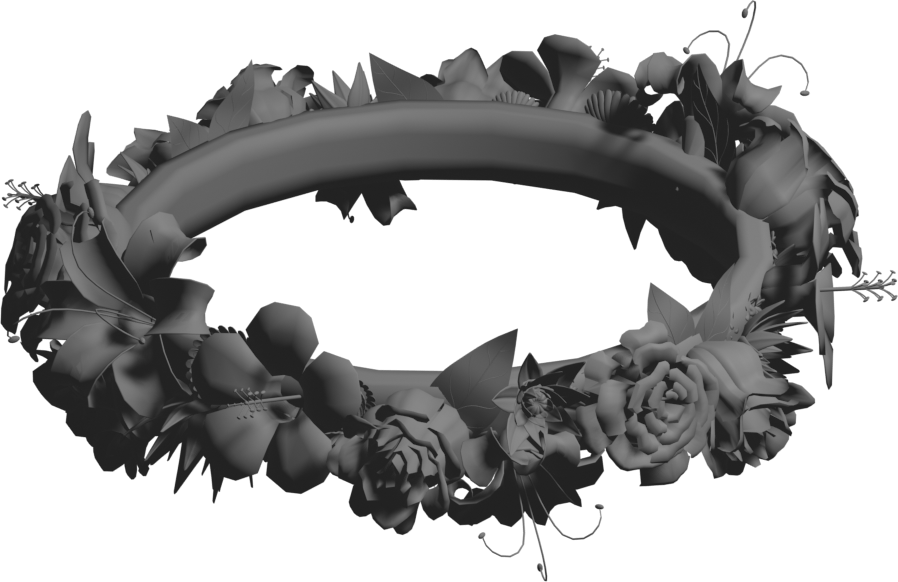}
    \caption{
        A rendering of the model ``Flower Crown'' (CC-BY) by the user ``Zoinks360''.
        The model is untextured.
    }
    \label{fig:no-texture}
\end{figure}

While the Sketchfab 3D Creative Commons Collection has been designed as a contemporary research collection usable for a broad range of tasks, the collection procedure has been selected in order to collect models useful for tasks for which existing collections are not sufficiently suitable.
In comparison to previously proposed collections, the S3D3C deliberately consists of a heterogeneous set of 3D models covering a wide variety of modern technical 3D model properties, such as different textures, materials, animations, and even some models with sound.
These varied technical properties enable evaluation of modern, view-based 3D model analysis and retrieval methods under more realistic and challenging scenarios.

\begin{figure}
    \centering
    \includegraphics[width=.35\linewidth]{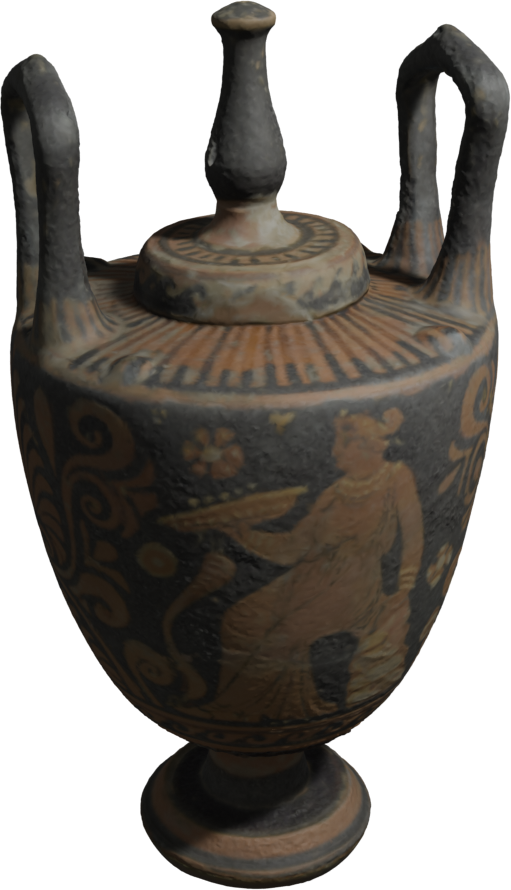}
    \caption{
        A rendering of the model ``Apulian Red-Figured Lebes Gamikos'' (CC0) by the user ``The Hunt Museum''.
        The model is textured entirely by the material without any texture files.
    }
    \label{fig:material-texture}
\end{figure}

One such challenging scenario for view-based 3D model methods is the interplay of textures, materials, and lighting when rendering a view of a 3D model.
While the majority of models in the collection use a combination of materials and texture files, some, such as the flower crown rendered in \autoref{fig:no-texture}, are completely untextured, while others, such as the cauldron rendered in \autoref{fig:material-texture}, are textured solely through their material.

\begin{figure}
    \centering
    \resizebox{\linewidth}{!}{
        \includegraphics[height=3cm]{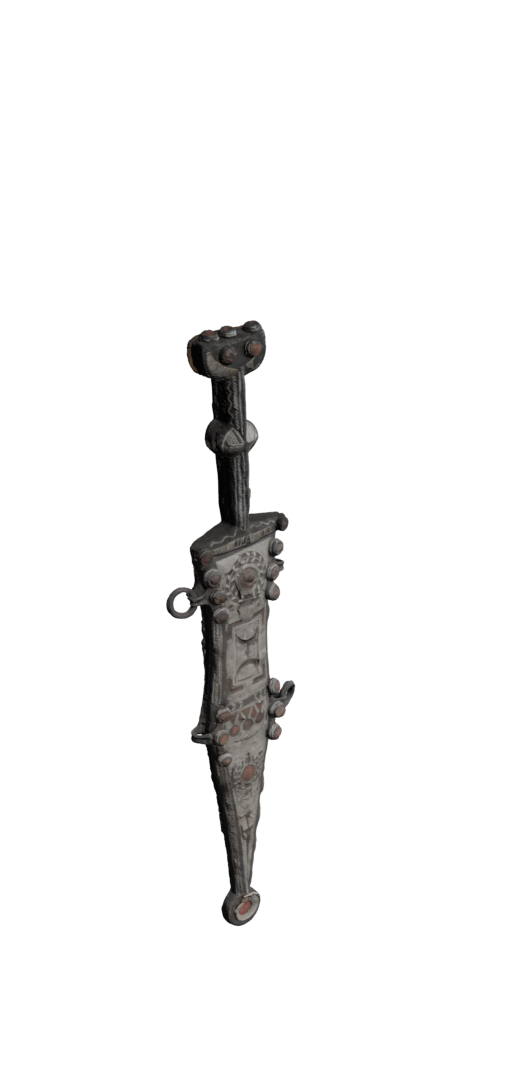}
        \includegraphics[height=3cm]{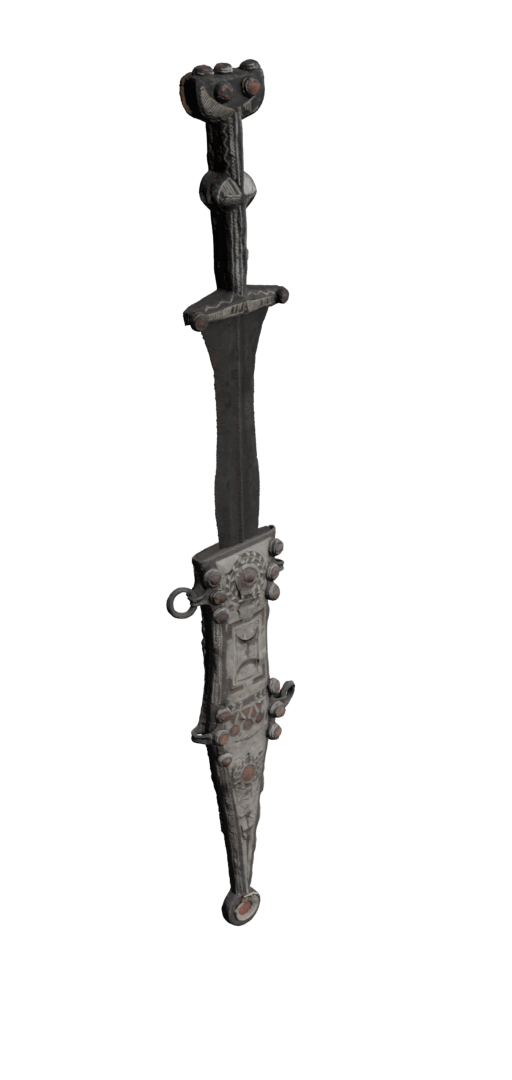}
        \includegraphics[height=3cm]{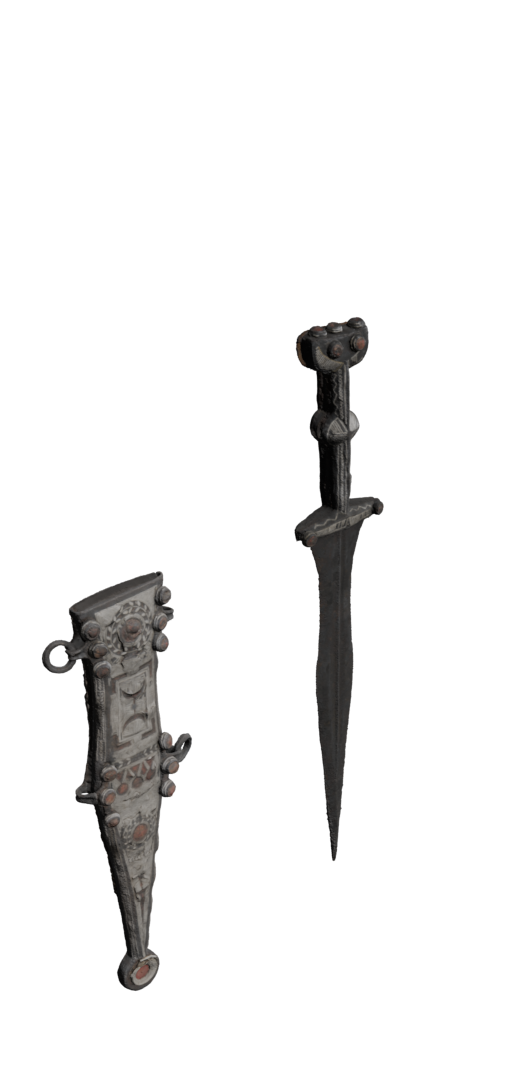}
    }
    \caption{
        Three frames from the animation of the model ``048 Dolch mit Scheide / Dagger with scabbard'' (CC0) by the user ``LWL-Archaeologie''.
        At the beginning of the animation the blade of the dagger is completely hidden by the scabbard. Over the course of the animation, the dagger is pulled from the scabbard and moved to the side.
    }
    \label{fig:animation}
\end{figure}

Animations are another challenge, which affect not only view-based but also geometry-based 3D model methods.
Animations may alter the geometry and the appearance of a model drastically over the course of each included animation, of which there may be several for the same model.
The number of animations in each model varies greatly, with the majority of animated models containing only a single animation, but others containing up to 169 animation tracks.
An example of how a model can change over the course of a relatively simple animation can be seen in \autoref{fig:animation}, in which a dagger is drawn from its scabbard.

\begin{figure*}
    \centering
    \resizebox{\linewidth}{!}{
        \includegraphics[height=3cm]{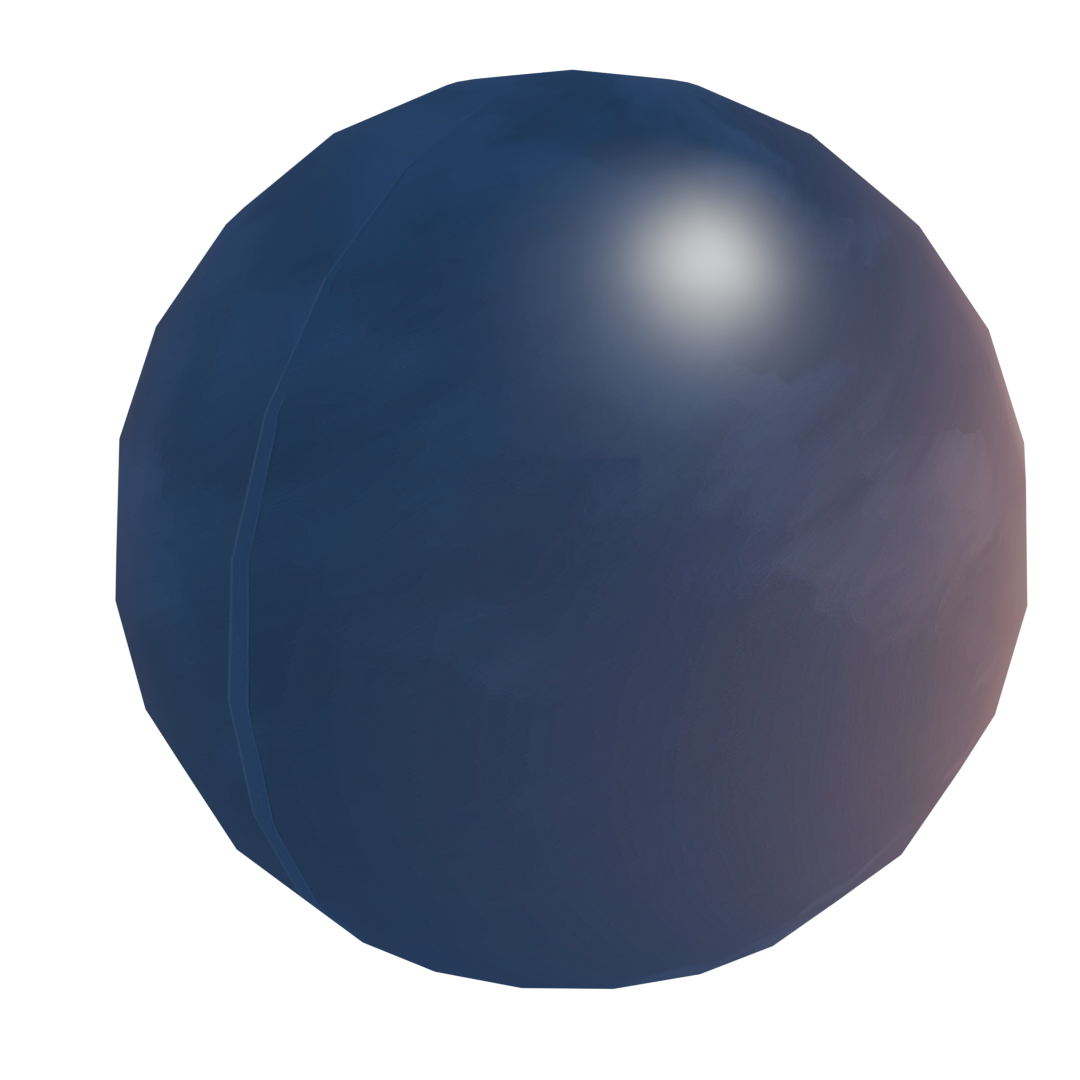}
        \includegraphics[height=3cm]{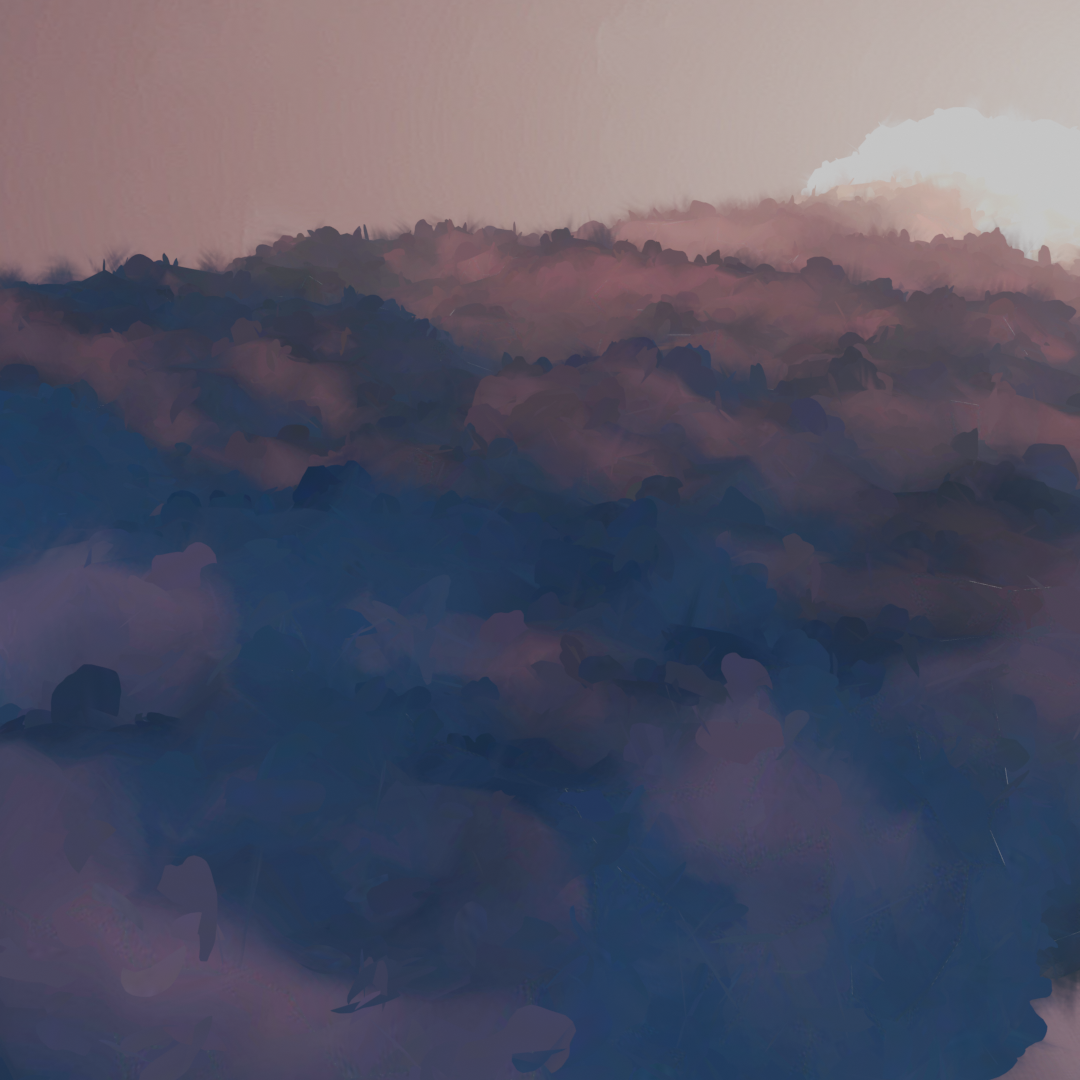}
        \includegraphics[height=3cm]{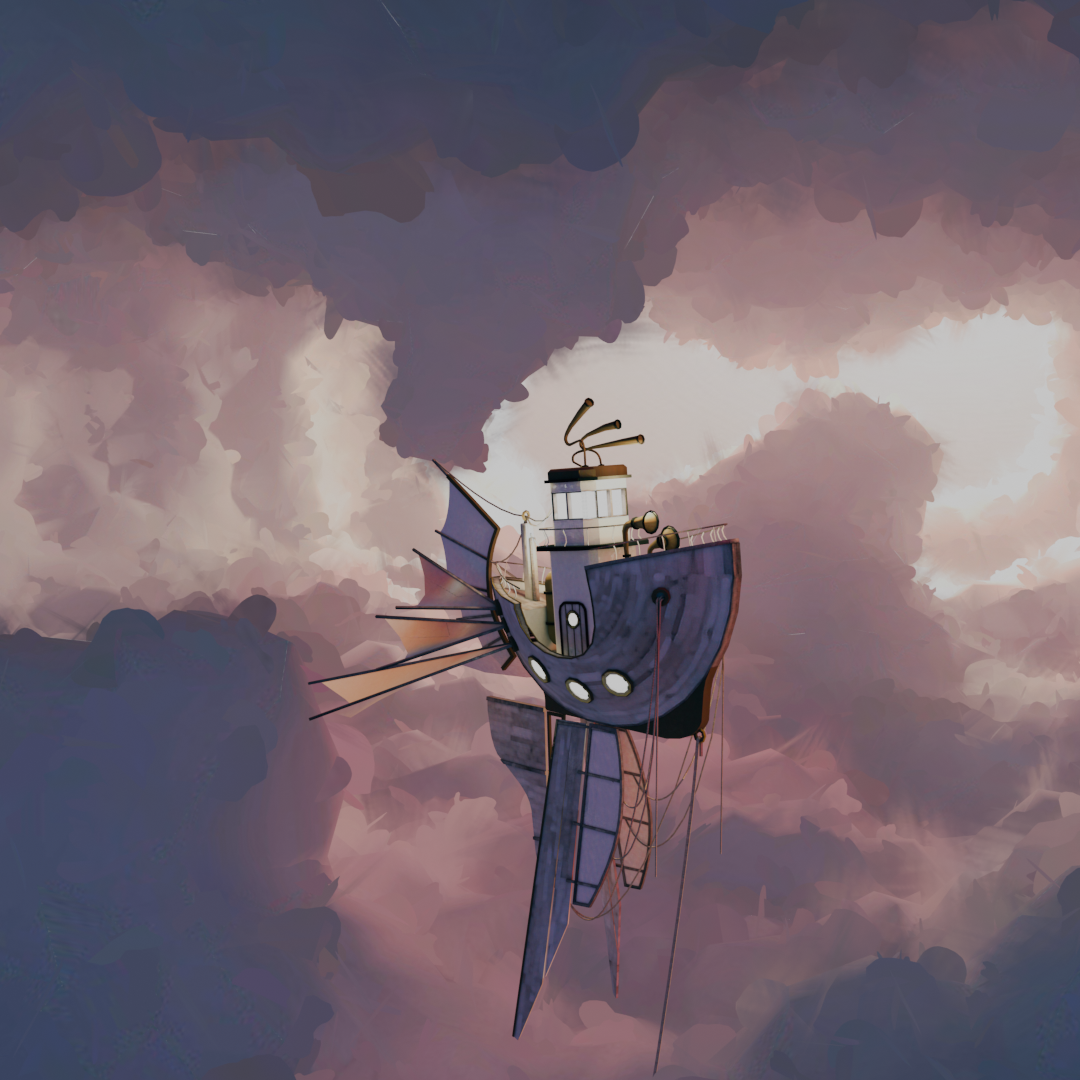}
        \includegraphics[height=3cm]{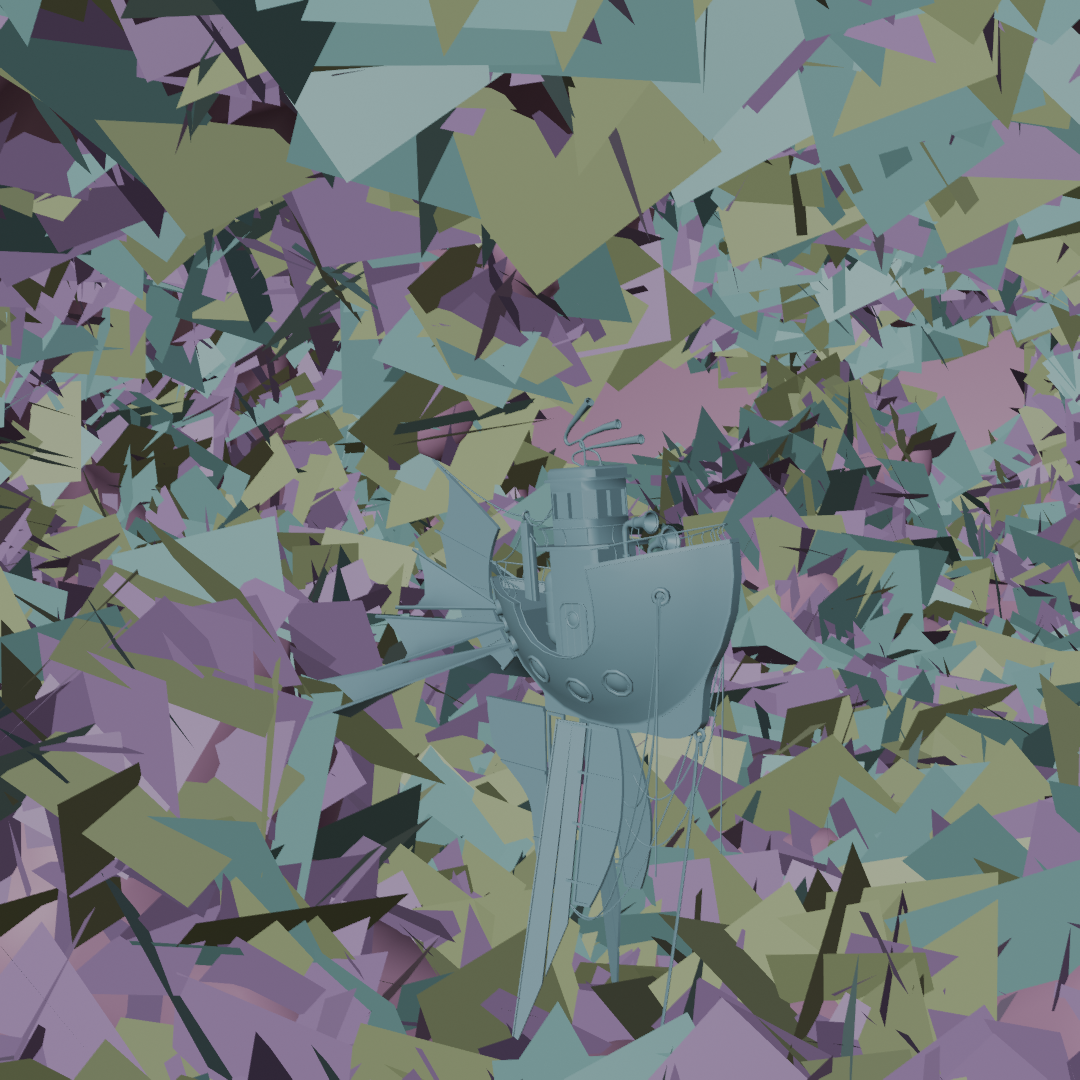}
    }
    \caption{
        Different views of the model ``Ship in Clouds'' (CC-BY) by the user ``Bastien Genbrugge''.
        From left to right: outside render of the model, render from inside the background sphere but outside the clouds, render from inside the clouds, and render of the raw geometry with random colors from inside the clouds.
    }
    \label{fig:different-views}
\end{figure*}

The S3D3C contains a mixture of digitally created models, such as already shown in \autoref{fig:no-texture}, and 3D scans, such as in \autoref{fig:material-texture} and \autoref{fig:animation}.
This mixture of model origin is important, as digitally created and 3D scanned models pose very different challenges to view- and geometry-based methods.
An especially interesting example of a digitally created 3D model can be seen rendered in four different ways in \autoref{fig:different-views}.
The model consists of a textured sphere as a background containing a layer of clouds, which in turn contains a detailed ship.
It is clear that the artists intent for the model is to be rendered from within the layer of clouds towards the ship.
In cases such as this, current automated view-based methods for 3D model analysis may struggle to find appropriate camera placements to properly capture the intended content of the model.
Geometry-based methods may also have difficulties identifying the main focus of the model, but will additionally struggle with the geometry of the layer of clouds, which can be seen to consist entirely of quadrilaterals in the render to the far right of \autoref{fig:different-views}.

\subsection{Limitations}

While the Sketchfab 3D Creative Commons Collection is a modern collection containing a variety of 3D models with technical properties representative of models in use today, it has some limitations.
Its limitations stem primarily from the way in which the models have been collected.

Although the number of models in the S3D3C is much larger than that of many existing collections, it is still relatively small in comparison to the number of 3D models in use today, and as such it may be difficult to evaluate methods meant to work on much larger collection sizes.
As of December 2022, Sketchfab reports that it is hosting more than 5 million 3D models~\cite{crawfordSketchfabCelebrates102022}.
While many of these models are not creative commons licensed and could therefore not appear in the S3D3C, only a limited number of the creative commons licensed models could be collected due to rate limits of the Sketchfab download API and page limits of the search API.

Certain categories of 3D model occur more often in the collection than others, such as cultural heritage 3D models.
This imbalance most likely comes from the selection of licenses collected in the S3D3C, as many public memory institutions, such as museums and archives, upload their 3D model scans to Sketchfab under creative commons licenses.
Additionally, due to restrictions on the distribution of certain kinds of 3D data, some categories, such as medical models, may be underrepresented in the S3D3C.

The S3D3C contains a lot of user created metadata thanks to model title, descriptions, tags, and categories provided by the users who uploaded the models.
While the availability of this metadata is a great advantage, the quality of this data varies greatly between models, with some containing long and detailed descriptions and others missing the description entirely.
The same applies to categories and tags, where not all models may be labelled with the same level of detail.

\section{Conclusion}
\label{sec:conclusion}

In this paper we present the Sketchfab 3D Creative Commons Collection (S3D3C), a new 3D model collection containing creative commons licensed models from the Sketchfab 3D model platform.
The collection contains models with a variety of different technical properties and metadata, which go beyond those included in previously proposed collections and enable the evaluation of state-of-the-art view- and geometry-based 3D model analysis and retrieval methods.
We make the collection available as a list of IDs for download directly from Sketchfab, and plan to make it available additionally as a direct download in order to ensure that models removed from Sketchfab remain accessible.

\begin{credits}
\subsubsection{\ackname} This work was partly supported by the Swiss National Science Foundation through project \href{https://data.snf.ch/grants/grant/193788}{``Participatory Knowledge Practices in Analog and Digital Image Archives''} (contract no.\ 193788) and the InnoSuisse project Archi\-panion (contract no.\ 2155012542).
\end{credits}
%
%
%
\bibliographystyle{splncs04}
\bibliography{bibliography}

\end{document}